\documentclass[12pt,a4paper]{article}
\usepackage[utf8]{inputenc}
\usepackage[T1]{fontenc}
\usepackage[top=2.5cm, bottom=2.5cm, left=2.0cm, right=2.0cm]{geometry}

\usepackage{amscd,amsfonts,amsmath,amssymb,amsthm,bbm,bm,latexsym,mathrsfs,mathtools,thmtools}
\usepackage[subnum]{cases}	
\allowdisplaybreaks			

\usepackage{hyperref} 		
\hypersetup{bookmarks=true, unicode=true, urlcolor=blue, linkcolor=blue, filecolor=blue, colorlinks=true, citecolor=black, final=true}

\usepackage{xcolor,subfigure,epsfig,graphics,graphicx}
\usepackage{caption}
\usepackage{enumitem}
\usepackage{longtable,float,array}
\usepackage{setspace}
\usepackage{threeparttable,booktabs,rotating}

\makeatother
\title{\textbf{Public Concern and the Financial Markets during the COVID-19 outbreak}}

\author{
\hspace*{-20pt}
Michele Costola\textsuperscript{a}\thanks{e-mail: \href{mailto: michele.costola@unive.it}{michele.costola@unive.it}}
\hspace*{12pt} 
Matteo Iacopini\textsuperscript{a}\thanks{e-mail: \href{mailto: matteo.iacopini@unive.it}{matteo.iacopini@unive.it}}
\hspace*{12pt} 
Carlo R.M.A. Santagiustina\textsuperscript{a}\thanks{e-mail: \href{mailto: carlo.santagiustina@unive.it}{carlo.santagiustina@unive.it}} \\ \\
\small \textsuperscript{a}Ca' Foscari University of Venice, Cannaregio 873, 30123 Venice, Italy
}

\date{\vspace*{10pt}\normalsize \today}

\begin{document}

\maketitle

\begin{abstract}
We measure public concern in Italy, Germany, France, Great Britain, Spain and the United States during the outbreak of COVID-19 using three search-engine data sources from Google Trends: YouTube, Google News and Google Search.
We find that the dynamic of public concern in Italy is a driver of that in other countries.
Among the Google trends series, we document that the Italian index is more relevant than the others in explaining stock index returns for all countries.
Finally, we perform a time-varying analysis and observe that the most severe impacts on the financial markets occur at each step of the Italian lock-down process.
\end{abstract}

\section{Introduction}
Google Trends (GT) provides indices based on the relative web-search volumes of a specific topic over time. These indices can be retrieved for selected geographic areas or at the worldwide scale. The interpretation of GT indices is straightforward: the higher the value of a given GT index, the greater the public attention for that topic.
In recent years, the informational content of Google Trends data has shown to have explanatory and forecasting power in several fields of economics and finance.
In relation to financial markets, GT indices based on stock-related terms: (i) lead to a higher portfolio diversification with respect to the market benchmark \cite{kristoufek2013}, and (ii) can be used as sentiments \cite{da2011search} and early warning market signals \cite{bijl2016,curme2014quantifying}.
On the macroeconomic side, GT indices have been used to construct economic uncertainty indicators able to explain several macroeconomic variables \cite{donadelli2015,Castelnuovo2017google,d2017predictive}.
GT data has also been successfully used for disease surveillance purposes for MERS \cite{shin2016high}, Chicken-Pox \cite{bakker2016digital} and Flu \cite{yang2015accurate}.

In this paper, we retrieve GT country indices (GT-COVID-19) for the coronavirus topic from January 2020 to April 2020. We use them as proxies for country-level public concern and investigate their impact on the financial markets during the outbreak of the coronavirus disease. The COVID-19 pandemic is an interesting case to investigate since, due to its virulence and infectivity, it represents a major exogenous shock to the economic and financial system, which could not have been reasonably foreseen.
In our analysis, we consider the six most impacted countries worldwide in terms of confirmed cases, as of May the 1st 2020: the United States, Spain, Italy, Great Britain, Germany and France.
Our findings are three-fold. First, the GT-COVID-19 index for Italy is found to be a driver of the GT indices for all the countries considered. This lead-lag relationship is primarily due to the fact that Italy has been the first European country to experience an outbreak of COVID-19 and to implement lock-down measures since World War II.
In addition, several European governments have introduced travel restrictions from and to Italy during the first weeks of the outbreak.
The ensuing spread of the coronavirus disease in other countries provided a similar dynamic on the new cases and for the implemented lock-down measures. Therefore, the delayed reaction of those indexes is probably due to the twist of the aforementioned events. 
Second, we analyse if GT-COVID-19 indices explain the stock market returns. Given that an epidemic disease is by definition an adverse event, GT indices can be interpreted as a measure of coronavirus-related uncertainty and perceived risk.

Our findings show that GT-COVID-19 indices contribute to explain the dynamic of the stock market returns for Italy, Spain and Germany.
Interestingly, substituting country-specific GT indices with the Italian one magnifies the exposition of all considered markets to COVID-related public concern, bursting the explained variance of stock index returns.

This highlights that the pandemic outbreak in Italy may have exerted a role in the general perception of the pandemic severity. In this respect, we finally perform a time-varying analysis to investigate the impact of the GT-COVID-19 indices on the financial markets over time. Interestingly, we identify the most severe impacts during each step of the Italian lock-down process and find almost no impact prior to the beginning of the pandemic in Europe, even though the presence of COVID-19 in China was already known by the end of December 2019.

The structure of the paper is as follows: Section~\ref{sec:data} briefly illustrates the dataset and data collecting process, then Section~\ref{sec:application} presents the main results. Finally, Section~\ref{sec:conclusion} concludes.

\section{Data}  \label{sec:data}
We collected social and financial data for a panel of $N=6$ countries, including Germany (DE), France (FR), United Kingdom (GB), Unites States (US), Italy (IT) and Spain (ES), for $T=75$ working days, from 1 January 2020 to 14 April 2020.

As for the financial data, we have downloaded the closing price of the stock market index\footnote{We used the following indices: S\&P500 (US), FTSE MIB 30 (Italy), DAX (Germany), CAC 40 (France), IBEX (Spain) and FTSE 100 (Great Britain).} at daily frequency from Bloomberg, and obtained the log-returns.

\paragraph{Social data collection}
The raw social data have been systematically collected from Google using the GtrendsR package by \cite{massicotte2016gtrendsr}, which sends requests to the Google Trends website
exploiting the query parameters in the URL of the request (see Table~\ref{tab:Trends}).

\begin{table}[!th] 
\centering
\renewcommand*{\arraystretch}{1.2}  
\footnotesize
\begin{tabular}{*{2}{|c}|*{1}{l}|}
\toprule 
 Param. & Value(s) & Description\\ 
\midrule 
\textbf{q}     & /m/01cpyy (\textit{coronavirus topic}) &  filters data by term(s) OR topic \\
\textbf{date}  & 01-01-2020 \textit{to} 14-04-2020 &  filters data through the time dimension \\
\textbf{geo}   & IT, FR, DE, US, GB, ES & filters by geographic area \\
\textbf{gprop} & (all), youtube, news & filters by source, defaults to (all) google searches \\
\bottomrule 
\end{tabular} 
\caption{\footnotesize Google Trends query parameters} 
\label{tab:Trends}
\end{table}


We have downloaded daily search relative volumes per country, from 01/01/2020 to 14/01/2020, on Google Search (all), Youtube and Google News, matching the ``coronavirus'' topic.
Then, the time series have been rescaled in the $[0,1]$ interval.
Topics are groups of terms that have equivalent meanings in different languages. Searching by topic, rather than by term(s), has several practical and theoretical advantages, including:
\begin{itemize}
    \item \textbf{No ``priors'' affecting results:} with this method a topic is given as it is, e.g. no means to strategically ``stretch it'' or ``prune it''. Researchers, on their side, have the advantage of not having to use their prior knowledge about the topic to choose the terms used to identify it. For example, one could erroneously specify a set of terms that are not representative for the topic he wants to study;
    \item \textbf{No need to ``translate'' the query terms:} the topic's identifier is \textit{unique for all languages} and all countries in the world. Therefore, with the same identifier one can extract data about a topic for any country, in whatever language. This is a huge advantage for studies that comprise different countries with different languages and different degree of usage of the various technical terms related to a topic (e.g COVID-19 \textit{vs} Coronavirus). It also allows to take into account searches done in a country in a language other than the official one(s)\footnote{E.g., see searches related to the word ``virus'' in Chinese, made from the US (\href{https://trends.google.com/trends/explore?date=2020-01-01\%202020-05-01&geo=US&q=\%E7\%97\%85\%E6\%AF\%92}{here}).}.
    Therefore the data obtained will also be representative of searches made by linguistic minorities/foreigners;
    \item \textbf{Overcoming the ``number of terms'' limit:} since Google Trends imposes a limit on the number of terms that can be used jointly to build queries, if a topic is related to many terms, using terms is never convenient.
\end{itemize}

\begin{figure}[!th]
\centering
\hspace*{-14pt}
\begin{tabular}{cccc}
\begin{rotate}{90} \hspace{30pt} $y_{i,t}$ \end{rotate} & \hspace{-12pt}
\includegraphics[trim= 28mm 0mm 24mm 0mm,clip,height= 3.0cm, width= 8.5cm]{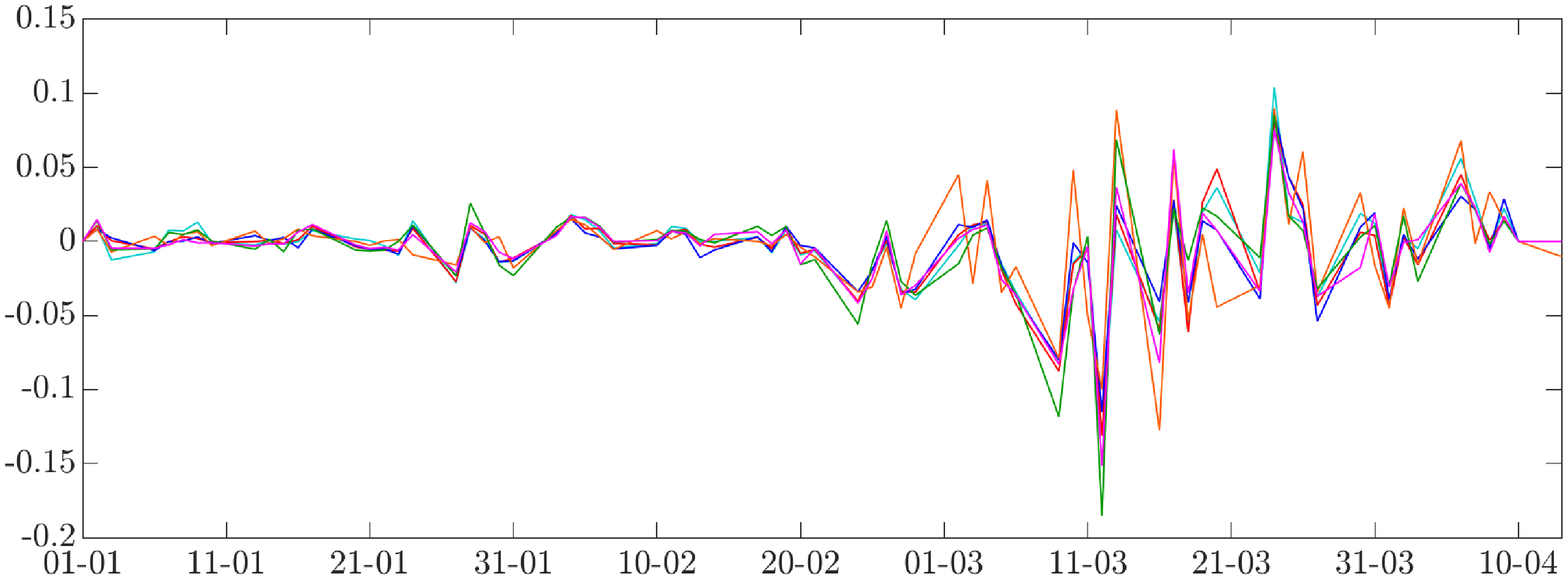} &
\begin{rotate}{90} \hspace{24pt} $GT_{N,i,t}$ \end{rotate} & \hspace{-12pt}
\includegraphics[trim= 28mm 0mm 24mm 0mm,clip,height= 3.0cm, width= 8.5cm]{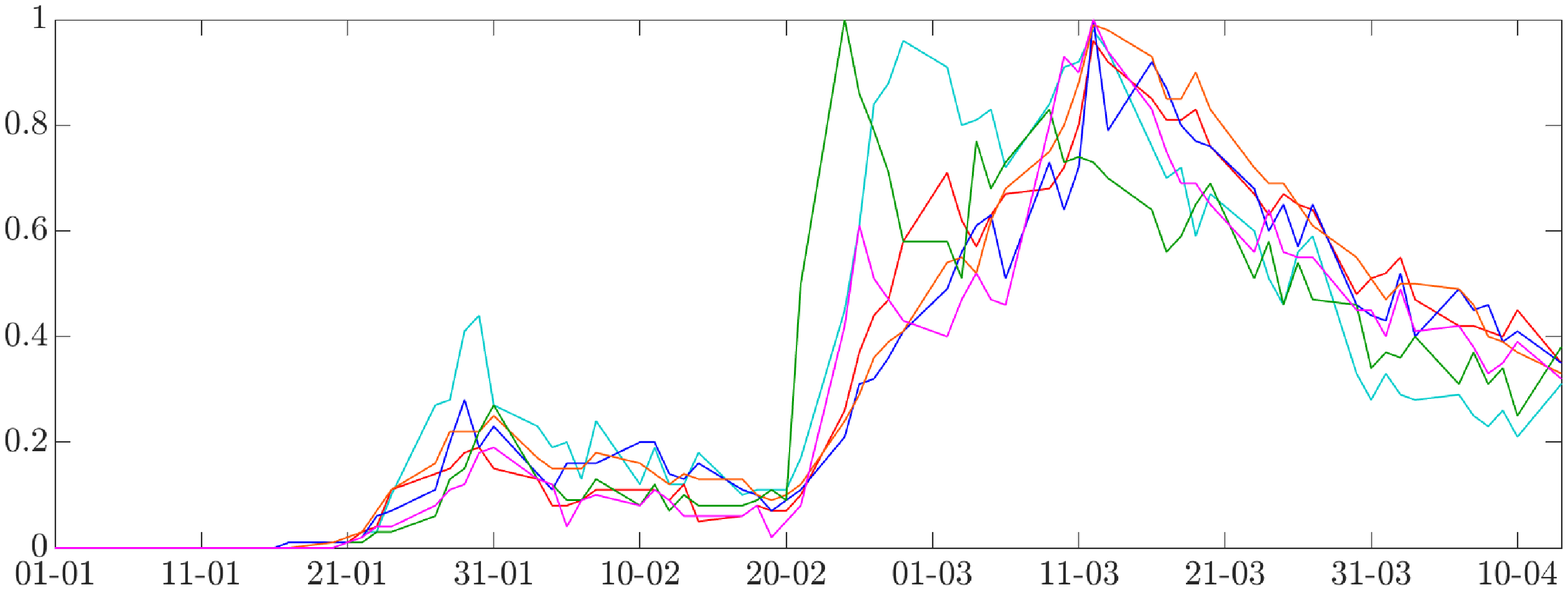} \\
\begin{rotate}{90} \hspace{24pt} $GT_{Y,i,t}$ \end{rotate} & \hspace{-12pt}
\includegraphics[trim= 28mm 0mm 24mm 5mm,clip,height= 3.0cm, width= 8.5cm]{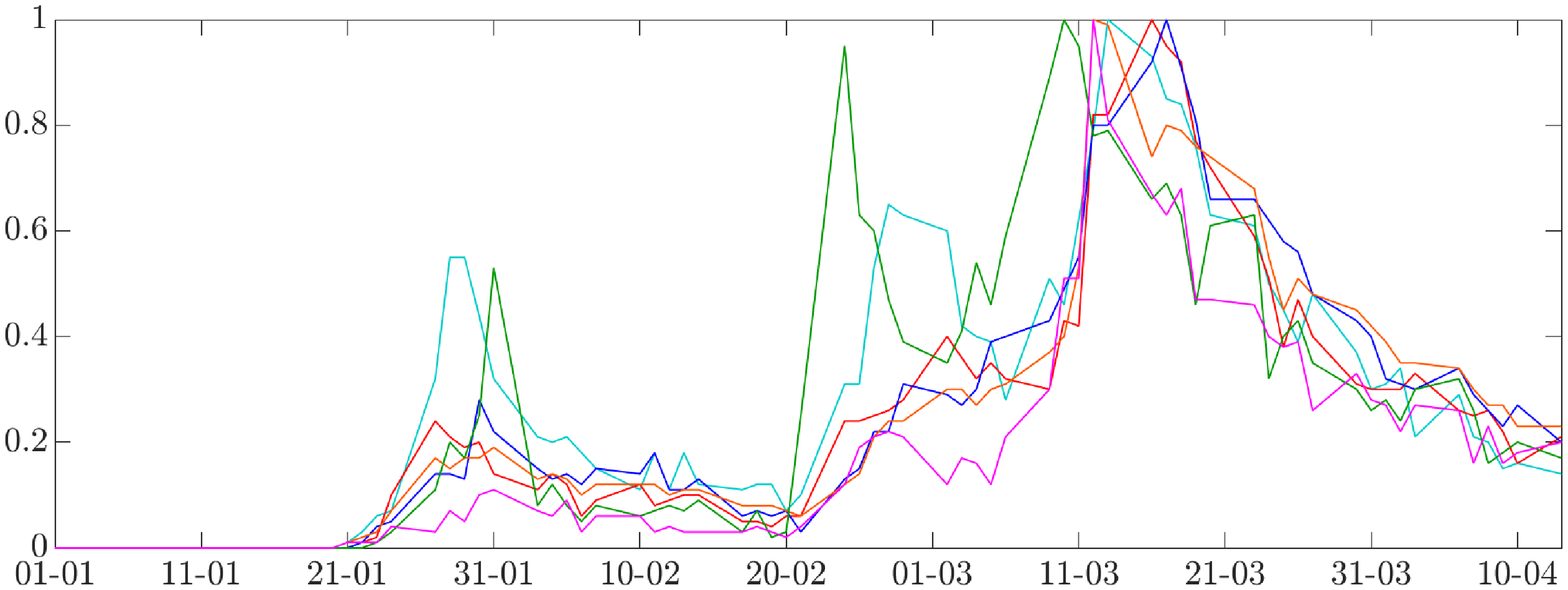} &
\begin{rotate}{90} \hspace{24pt} $GT_{S,i,t}$ \end{rotate} & \hspace{-12pt}
\includegraphics[trim= 28mm 0mm 24mm 0mm,clip,height= 3.0cm, width= 8.5cm]{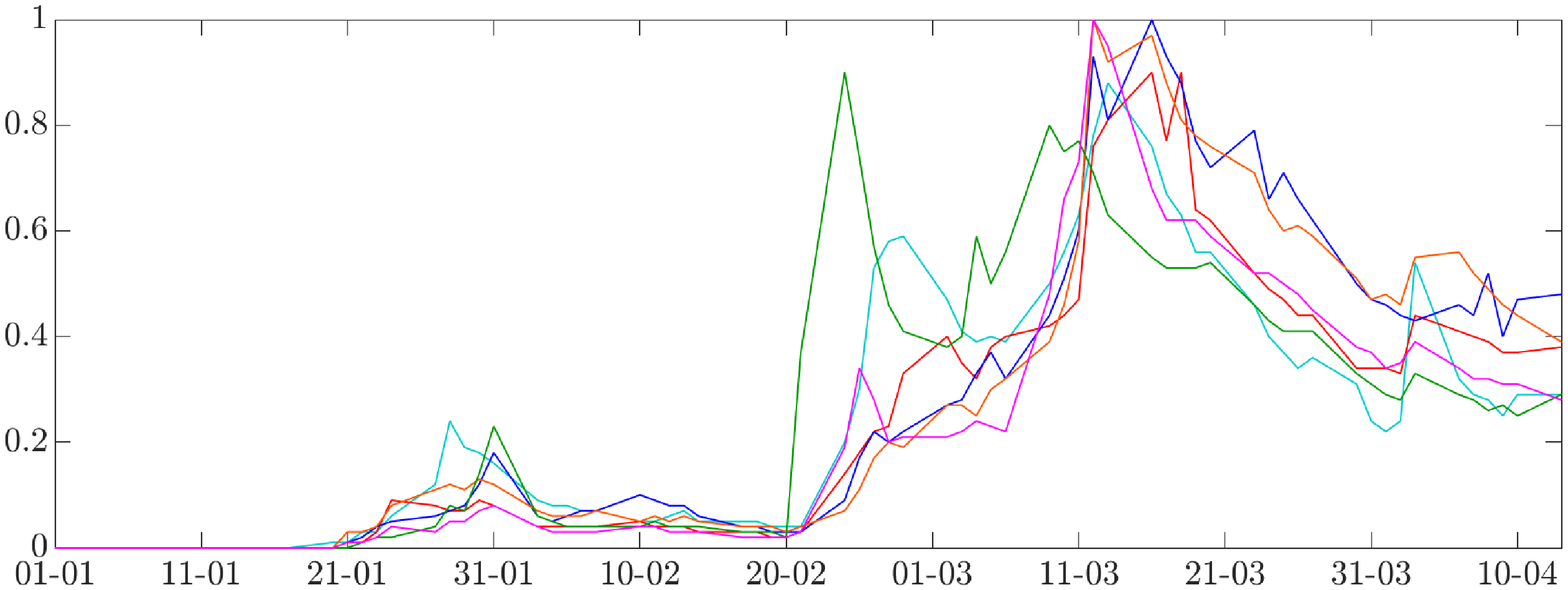}
\end{tabular}
\captionsetup{width=0.95\linewidth}
\caption{Sample: from 1 January 2020 to 14 April 2020 (financial market calendar).
Log-returns of market index (top-left row) and GT-COVID-19 series in levels for: Google News (top-right), YouTube (bottom-left) and Google Search (bottom-right).
Countries: DE (cyan), FR (red), GB (blue), US (orange), IT (green), ES (purple).}
\label{fig:raw_data}
\end{figure}

\paragraph{Preliminary analysis}
Figure~\ref{fig:raw_data} shows the time series of log-returns and the three groups of GT indices, subdivided by source.
For each country $i \in \{ DE, FR, GB, US, IT, ES \}$, we observe the log-return series, $y_{i,t}$, and three GT indices $GT_{j,i,t}$, where $j \in \{ Y, N, S \}$, corresponding to YouTube, Google News and Google Search, respectively.
The country-specific GT series have a common dynamic, whereas for each source there is cross-country heterogeneity.


%

We remark that high values of either GT index correspond to intervals which exhibit turbulent dynamics for financial returns. This is particularly evident throughout March.
Moreover, we notice that the Italian GT indices (for each source) anticipate those of the other countries. To test for the presence of lead-lag relations, we perform a cross-correlation analysis of each series against Italy. Table~\ref{tab:lead_lag} reports the peak in the cross correlation, where negative values implying that the series for Italy is leading.
By looking at the beginning of the period, we observe that both the GT indices and the financial series do not appear to react to epidemics-related facts originated in China.
Overall, these findings corroborate the interpretation that public concern has cross-country similarities, with Italy acting as the forerunner.

\begin{table}[H]
\centering
\begin{tabular}{c|ccccc}
              & DE & FR & GB & US & ES \\
\hline
YouTube       & -3 & -4 & -6 & -4 & -3 \\
Google News   & -3 & -5 & -7 & -5 & -3 \\
Google Search & -4 & -6 & -8 & -6 & -3 \\
\hline
\end{tabular}
\caption{Peak of the daily cross correlation of the GT-COVID-19 indices (in levels), each country series versus Italy. Negative values for a country series mean that Italy's corresponding series for is leading (conversely for positive values).}
\label{tab:lead_lag}
\end{table}

%
%

\section{Empirical results}   \label{sec:application}

\subsection{Constant parameters regression}
In this section we present the results of country-specific regressions, aiming to discover a possible relation between social media data and the returns on each country's market index.
We focus on YouTube data, that is $GT_{i,t} = GT_{Y,i,t}$, however we have obtained similar results using either Google News or Google Search data.\footnote{As a robustness check, we also run a regression using as covariate the first principal component estimated from the three series and obtained similar results. Results are available upon request to the authors.}

We investigate the exposition of markets to public concern, as proxied by GT-COVID-19 indices, by estimating the following AR(1)-X model. We focus on YouTube indices, that is $GT_{j,i,t} = GT_{Y,i,t}$. Analogous results have been obtained using either Google News or Google Search indices.
\begin{equation}
y_{i,t} = \alpha_i + \beta_i GT_{i,t} + \delta_i y_{i,t-1} + \epsilon_{i,t}.
\label{eq:reg_lag_on_IT}
\end{equation}



\begin{table}[H] 
\centering 
\small
\begin{tabular}{|l|*{6}{c}|} 
\toprule 
& DE& FR& GB& US& IT& ES\\ 
\midrule 
const & 0.006** & 0.002 & -0.000 & 0.005 & 0.011*** & 0.004 \\ 
& (0.003) & (0.003) & (0.003) & (0.004) & (0.004) & (0.004) \\ 
$GT_t$& -0.032** & -0.024 & -0.014 & -0.030* & -0.058*** & -0.050* \\ 
& (0.017) & (0.021) & (0.017) & (0.021) & (0.021) & (0.033) \\ 
$y_{t-1}$& -0.078 & -0.059 & -0.070 & -0.469*** & -0.309* & -0.225 \\ 
& (0.139) & (0.166) & (0.159) & (0.171) & (0.189) & (0.224) \\ 
& & & & & & \\ 
\midrule 
$\bar{R}^2$  & 0.047  & 0.012  & -0.007  & 0.200  & 0.189  & 0.107 \\ 
\bottomrule 
\end{tabular} 
\caption{\footnotesize Estimates of model \eqref{eq:reg_lag_on_IT} for all countries. Social data: YouTube (country-specific).\\
{\scriptsize Robust standard errors are in parentheses, *, ** and *** denote statistical significance at 10\%, 5\% and 1\%, respectively.}}
\label{tab:reg_lag_own_YoTu}
\end{table}

Table~\ref{tab:reg_lag_own_YoTu} highlights that the impact of the country-specific GT-COVID-19 index is significant and negative for all countries except France and Great Britain.
This implies that markets negatively respond to public concern, as proxied by GT-COVID-19 indices.

Following the insights from the lead-lag relations, we check the exposition of markets to public concern in Italy by estimating the model using $GT_{i,t} = GT_{IT,t}$ for all $i$.
Interestingly, from Table~\ref{tab:reg_lag_on_IT_YoTu}, we observe that the Italian GT-COVID-19 index is a key explanatory variable for all country-level stock index returns, and its use remarkably increases the adjusted $R^2$ as compared to using country-specific GT indices.
In particular, the markets in FR and GB are insensitive to the public concern proxied by their GT indices, but are negatively related to the Italian GT-COVID-19.
This highlights that the severity of the outbreak perceived from Italy represents a timely indicator of the destabilizing effect of the pandemic on  financial markets.

As robustness check, we have controlled for country-specific market implied volatility and growth rate of confirmed COVID-19 cases, our findings remain unchanged.



\begin{table}[H] 
\centering 
\small 
\begin{tabular}{|l|*{6}{c}|} 
\toprule 
& DE& FR& GB& US& IT& ES\\ 
\midrule 
const & 0.006** & 0.002 & -0.000 & 0.005 & 0.011*** & 0.004 \\ 
& (0.003) & (0.003) & (0.003) & (0.004) & (0.004) & (0.004) \\ 
$GT_{IT,t}$& -0.032** & -0.024 & -0.014 & -0.030* & -0.058*** & -0.050* \\ 
& (0.017) & (0.021) & (0.017) & (0.021) & (0.021) & (0.033) \\ 
$y_{t-1}$& -0.078 & -0.059 & -0.070 & -0.469*** & -0.309* & -0.225 \\ 
& (0.139) & (0.166) & (0.159) & (0.171) & (0.189) & (0.224) \\ 
& & & & & & \\ 
\midrule 
$\bar{R}^2$  & 0.047  & 0.012  & -0.007  & 0.200  & 0.189  & 0.107 \\ 
\bottomrule 
\end{tabular} 
\caption{\footnotesize Estimates of model \eqref{eq:reg_lag_on_IT} for all countries. Social data: YouTube (country-specific).\\
{\scriptsize Robust standard errors are in parentheses, *, ** and *** denote statistical significance at 10\%, 5\% and 1\%, respectively.}}
\label{tab:reg_lag_on_IT_YoTu} 
\end{table}

\subsection{TVP analysis}
To further deepen the relation between the GT-COVID-19 indices and market returns, we estimate a time-varying parameter model, where the impact of the GT-COVID-19 index is allowed to vary according to a latent AR(1) process
\begin{equation}
\begin{split}
y_{i,t} & = \alpha_i + \beta_{i,t} GT_{i,t} + \delta_i y_{i,t-1} + \epsilon_{i,t}   \hspace*{20pt}  \epsilon_{i,t} \sim \mathcal{N}(0,\sigma_i) \\
\beta_{i,t} & = A_i \beta_{i,t-1} + \eta_{i,t}   \hspace*{93pt}  \eta_{i,t} \sim \mathcal{N}(0,B_i),
\end{split}
\end{equation}
where $\epsilon_{i,t}$ and $\eta_{i,t}$ are mutually independent. Estimation of $\beta_{i,t}$ is performed using the Kalman smoother, without imposing stationarity restrictions on the autoregressive coefficients $A_i$.

\begin{figure}[H]
\centering
\hspace*{-20pt}
\begin{tabular}{c c c c}
\begin{rotate}{90} \hspace*{30pt} \small DE \end{rotate} & \hspace*{-15pt}
\includegraphics[trim= 24mm 0mm 18mm 0mm,clip,height= 3.0cm, width= 8.7cm]{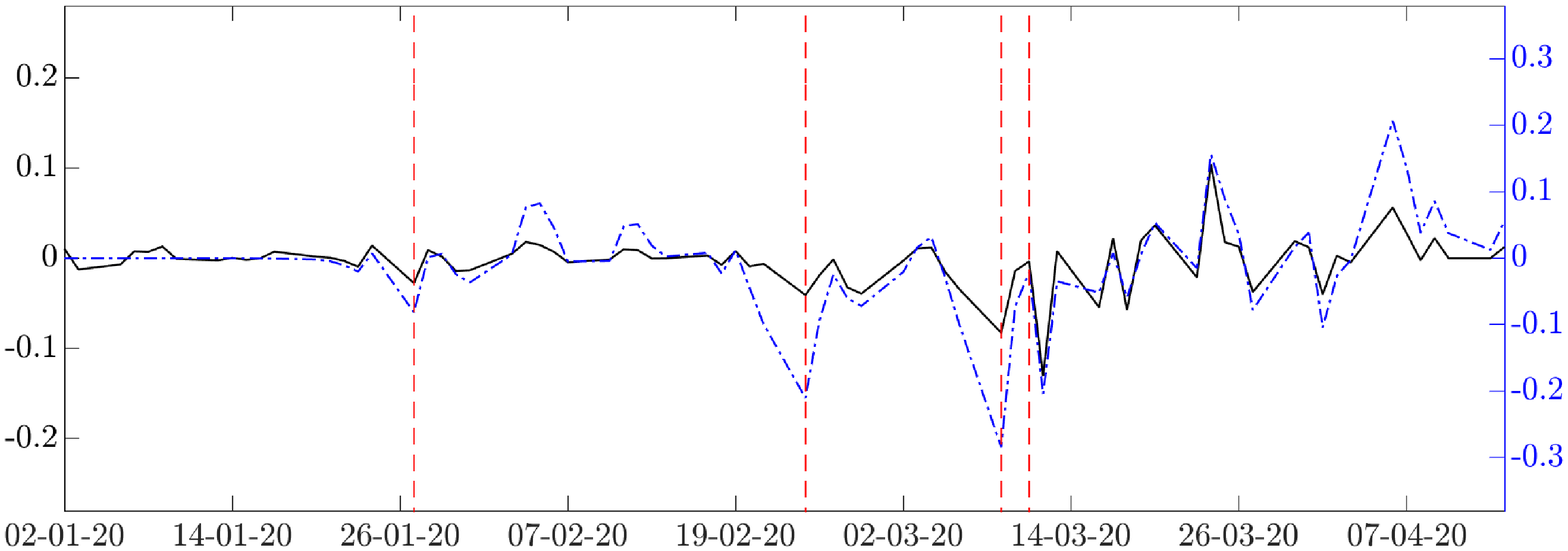} &
\begin{rotate}{90} \hspace*{30pt} \small FR \end{rotate} & \hspace*{-15pt}
\includegraphics[trim= 24mm 0mm 18mm 0mm,clip,height= 3.0cm, width= 8.7cm]{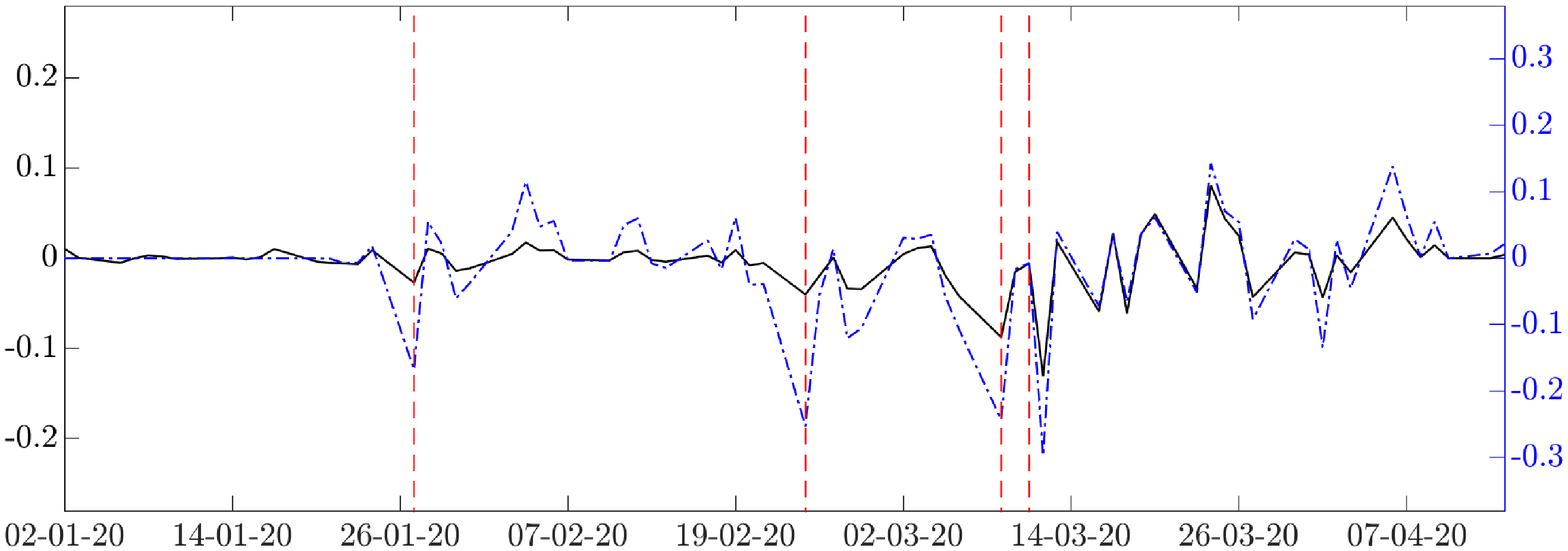} \\
\begin{rotate}{90} \hspace*{30pt} \small GB \end{rotate} & \hspace*{-15pt}
\includegraphics[trim= 24mm 0mm 18mm 0mm,clip,height= 3.0cm, width= 8.7cm]{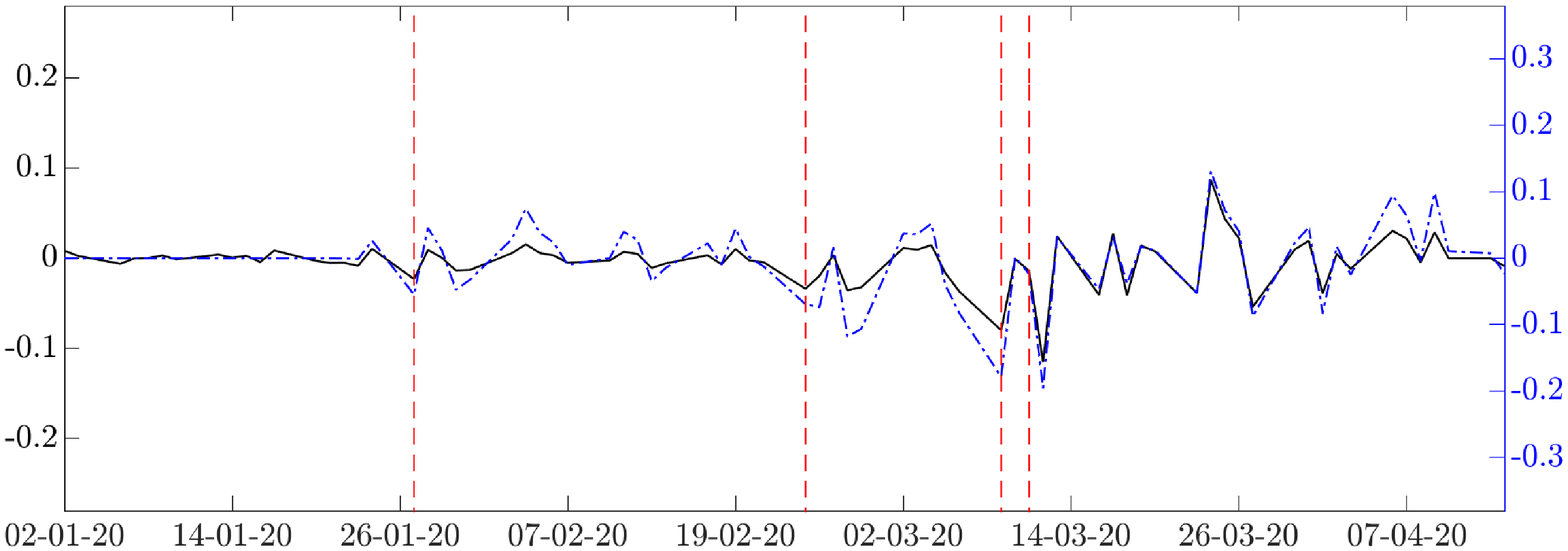} &
\begin{rotate}{90} \hspace*{30pt} \small US \end{rotate} & \hspace*{-15pt}
\includegraphics[trim= 24mm 0mm 18mm 0mm,clip,height= 3.0cm, width= 8.7cm]{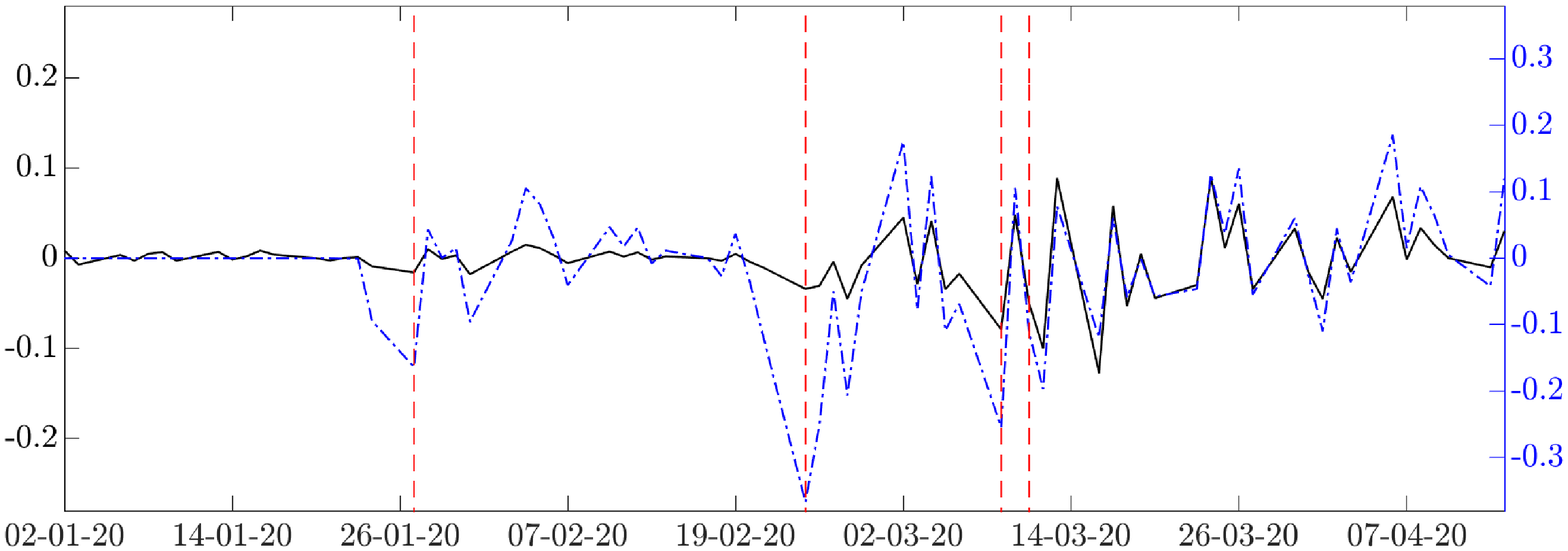} \\
\begin{rotate}{90} \hspace*{30pt} \small IT \end{rotate} & \hspace*{-15pt}
\includegraphics[trim= 24mm 0mm 18mm 0mm,clip,height= 3.0cm, width= 8.7cm]{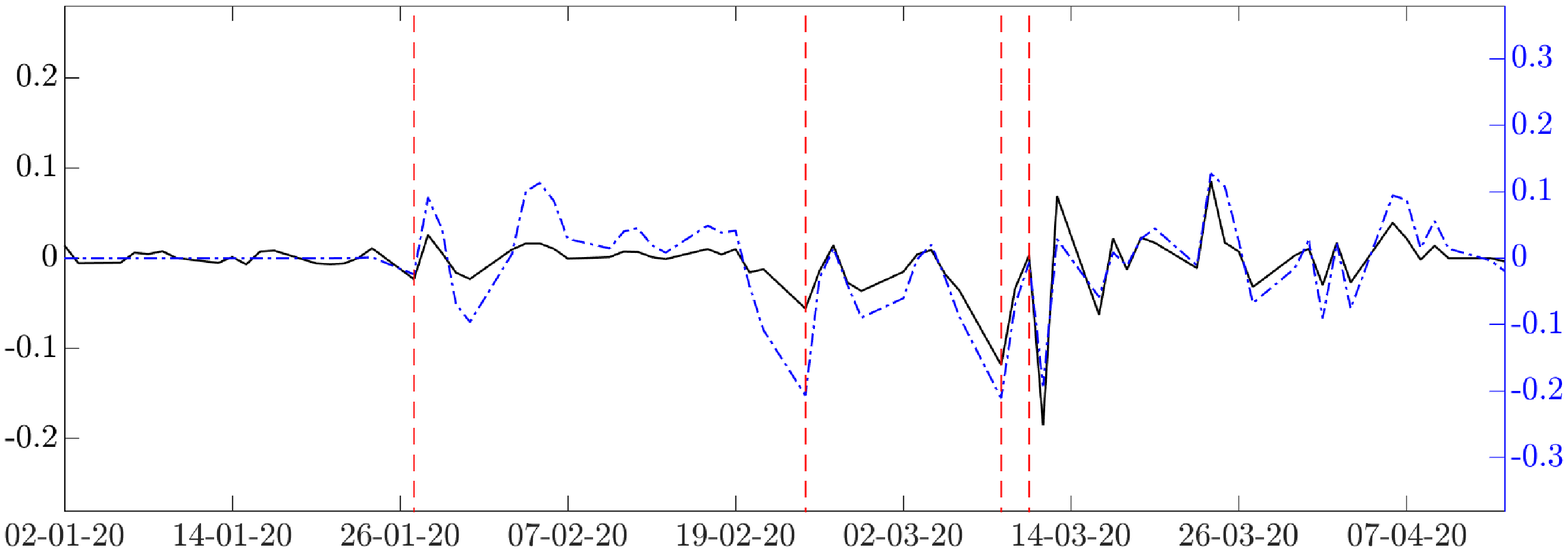} &
\begin{rotate}{90} \hspace*{30pt} \small ES \end{rotate} & \hspace*{-15pt}
\includegraphics[trim= 24mm 0mm 18mm 0mm,clip,height= 3.0cm, width= 8.7cm]{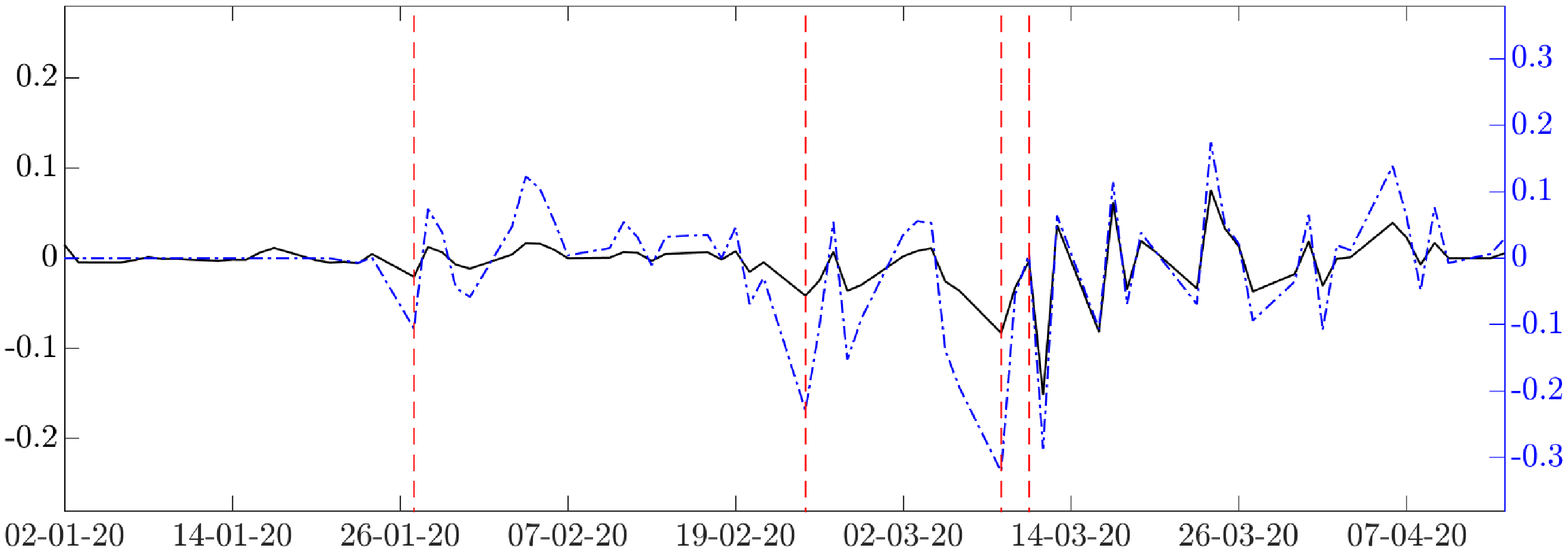}
\end{tabular}
\captionsetup{width=0.9\linewidth}
\caption{Log-returns (black, solid line) and estimated coefficient $\hat{\beta}_{i,t}$ (blue, dot-dashed line) of GT-COVID-19 index for YouTube.
Vertical red lines correspond to 23 January (first COVID-19 case in Germany), 24 February (lock-down of northern Italian provinces), 9 March (lock-down for all Italian citizens), 12 March (lock-down of most Italian economic activities).}
\label{fig:TVP_model}
\end{figure}

Figure~\ref{fig:TVP_model} plots the estimated paths of the impact of the GT index on the corresponding stock index returns.
All country-specific coefficients share a common dynamic.
The impact of all GT indices drops to negative values the same day of key events related to COVID-19. In particular, the lowest peaks occur in correspondence of the enactment of the Italian lock-downs.
Moreover, before the first confirmed case in Europe (23 January) all the coefficients are very close to zero. This signals that the GT indices had no explanatory power prior to the spreading of the pandemic to Europe, even though the presence of COVID-19 in China was already known since the end of December 2019.

\paragraph{Robustness checks}
The main findings of the constant parameter analysis are robust to controlling for implied volatility and number of confirmed cases (see Appendix~\ref{AppendixA}).
In addition, we have checked the robustness: (i) using a lagged version of the social index, (ii) inclusion of daily and weekly lags of the social index, controlling or not for the lagged financial return.\footnote{We refer to the Supplement for all these checks.}
Moreover, the results of the TVP analysis are robust to: (i) using a lagged version of the social index, (ii) controlling or not for the lagged financial return, (iii) specifying more lags in the dynamics of the coefficients.\footnote{The results are available upon request to the authors.}

\section{Conclusions} \label{sec:conclusion}
Recent literature has shown that Google Trends data can successfully explain the current and future patterns of the state of the economy, especially during unfavorable events \cite{heiberger2015collective,yu2019online,zhong2019revisiting}.
Differently from recent financial crises, COVID-19 is an exogenous shock to the system that has affected several countries with different timings related to the spreading of the disease.
We have investigated the exposure of the stock index returns of Italy, France, Germany, Great Britain, the United States and Spain to GT-COVID-19 indices based on search-engine query volumes.

Our findings show that most of these indices have significant explanatory power on stock market returns. 
Interestingly, the Italian GT-COVID-19 index acts as forerunner and better explains other countries' market returns. 
Moreover, the greatest impact of GT indices occur in correspondence of the different phases of the lock-down in Italy, despite the public awareness of the contagion in China since January.
The disruptive effect of COVID-19 on financial markets is well described by the public concern perceived in Italy, which has been the first Western country to experience a virulent outbreak and to adopt drastic measures after China.

%

\bibliographystyle{plain} 
\bibliography{biblio}

\begin{thebibliography}{1}

\bibitem{bijl2016}
Laurens Bijl, Glenn Kringhaug, Peter Moln{\'a}r, and Eirik Sandvik.
\newblock Google searches and stock returns.
\newblock {\em International Review of Financial Analysis}, 45:150--156, 2016.

\bibitem{Castelnuovo2017google}
Efrem Castelnuovo and Trung~Duc Tran.
\newblock Google it up! a {Google} trends-based uncertainty index for the
  united states and australia.
\newblock {\em Economics Letters}, 161:149--153, 2017.

\bibitem{donadelli2015}
Michael Donadelli.
\newblock Google search-based metrics, policy-related uncertainty and
  macroeconomic conditions.
\newblock {\em Applied Economics Letters}, 22(10):801--807, 2015.

\bibitem{kristoufek2013}
Ladislav Kristoufek.
\newblock Can google trends search queries contribute to risk diversification?
\newblock {\em Scientific reports}, 3:2713, 2013.

\bibitem{massicotte2016}
Philippe Massicotte and Dirk Eddelbuettel.
\newblock gtrendsr: Perform and display google trends queries.
\newblock {\em R package version}, 1(3):5, 2016.

\bibitem{preis2013}
Tobias Preis, Helen~Susannah Moat, and H~Eugene Stanley.
\newblock Quantifying trading behavior in financial markets using google
  trends.
\newblock {\em Scientific reports}, 3:1684, 2013.

\end{thebibliography}


\begin{thebibliography}{10}

\bibitem{bakker2016digital}
Kevin~M Bakker, Micaela~Elvira Martinez-Bakker, Barbara Helm, and Tyler~J
  Stevenson.
\newblock Digital epidemiology reveals global childhood disease seasonality and
  the effects of immunization.
\newblock {\em Proceedings of the National Academy of Sciences},
  113(24):6689--6694, 2016.

\bibitem{bijl2016}
Laurens Bijl, Glenn Kringhaug, Peter Moln{\'a}r, and Eirik Sandvik.
\newblock Google searches and stock returns.
\newblock {\em International Review of Financial Analysis}, 45:150--156, 2016.

\bibitem{Castelnuovo2017google}
Efrem Castelnuovo and Trung~Duc Tran.
\newblock Google it up! a {Google Trends}-based uncertainty index for the
  {United States} and {Australia}.
\newblock {\em Economics Letters}, 161:149--153, 2017.

\bibitem{curme2014quantifying}
Chester Curme, Tobias Preis, H~Eugene Stanley, and Helen~Susannah Moat.
\newblock Quantifying the semantics of search behavior before stock market
  moves.
\newblock {\em Proceedings of the National Academy of Sciences},
  111(32):11600--11605, 2014.

\bibitem{da2011search}
Zhi Da, Joseph Engelberg, and Pengjie Gao.
\newblock In {S}earch of {A}ttention.
\newblock {\em The Journal of Finance}, 66(5):1461--1499, 2011.

\bibitem{donadelli2015}
Michael Donadelli.
\newblock Google search-based metrics, policy-related uncertainty and
  macroeconomic conditions.
\newblock {\em Applied Economics Letters}, 22(10):801--807, 2015.

\bibitem{d2017predictive}
Francesco D’Amuri and Juri Marcucci.
\newblock The predictive power of {Google} searches in forecasting {US}
  unemployment.
\newblock {\em International Journal of Forecasting}, 33(4):801--816, 2017.

\bibitem{heiberger2015collective}
Raphael~H Heiberger.
\newblock Collective attention and stock prices: {E}vidence from {Google
  Trends} data on {Standard and Poor's} 100.
\newblock {\em PloS one}, 10(8):e0135311, 2015.

\bibitem{kristoufek2013}
Ladislav Kristoufek.
\newblock Can {Google Trends} search queries contribute to risk
  diversification?
\newblock {\em Scientific reports}, 3:2713, 2013.

\bibitem{massicotte2016gtrendsr}
Philippe Massicotte and Dirk Eddelbuettel.
\newblock gtrendsr: {P}erform and display {Google Trends} queries.
\newblock {\em R package version}, 1(3):5, 2016.

\bibitem{shin2016high}
Soo-Yong Shin, Dong-Woo Seo, Jisun An, Haewoon Kwak, Sung-Han Kim, Jin Gwack,
  and Min-Woo Jo.
\newblock High correlation of {Middle East} respiratory syndrome spread with
  {Google} search and {Twitter} trends in {Korea}.
\newblock {\em Scientific reports}, 6:32920, 2016.

\bibitem{yang2015accurate}
Shihao Yang, Mauricio Santillana, and Samuel~C Kou.
\newblock Accurate estimation of influenza epidemics using {Google} search data
  via {ARGO}.
\newblock {\em Proceedings of the National Academy of Sciences},
  112(47):14473--14478, 2015.

\bibitem{yu2019online}
Lean Yu, Yaqing Zhao, Ling Tang, and Zebin Yang.
\newblock Online big data-driven oil consumption forecasting with {G}oogle
  {T}rends.
\newblock {\em International Journal of Forecasting}, 35(1):213--223, 2019.

\bibitem{zhong2019revisiting}
Xu~Zhong and Michael Raghib.
\newblock Revisiting the use of {W}eb search data for stock market movements.
\newblock {\em Scientific reports}, 9(1):1--8, 2019.

\end{thebibliography}

\clearpage
\appendix

\section{Data collection}
Google Trends provides a rescaled values of relative search volumes per  unit-of-time. The data-generation procedure can summarized as follows:
\begin{enumerate}
\item Extract a random sample of queries corresponding to the searched geographical area, category and time-span;
\item For every  unit-of-time (whose length/duration depends on the time-span) divide the count of the number of searches matching the query term(s)/topic parameter (\textbf{q}), by the total number of searches in the same unit-of-time;
\item Rescale the obtained time series in the $[0,100]$ interval by: (a) removing the minimum value of the series to all values, (b) dividing all values of the series by the maximum value of the series, (c) multiply all values by 100, (d) round all values to the nearest integer.
\end{enumerate}


\section{Robustness checks}  \label{AppendixA}
We checked the robustness of our findings by estimating the constant parameters model with additional lags of the GT indices, aiming to capture the daily and weekly effects.
Moreover, we controlled for (i) the country-specific implied volatility\footnote{VIX (US), FTSEMIB Index FLDS Moneyness (Italy), V1X (Germany), VCAC 40 (France), VIBEX (Spain) and IVIUK (Great Britain). Source: \textit{Bloomberg.}} ($IV_{i,t}$), and (ii) the country-specific growth rate of new COVID-19 cases\footnote{NCOVUSCA (US), NCOVITCA (Italy), NCOVDECA (Germany), NCOVFRCA (France), NCOVESCA (Spain) and NCOVUKCA (Great Britain). Source: \textit{Bloomberg.}} ($\Delta^{\%} CC_{i,t}$).

In the latter case, we estimated the model
\begin{equation*}
y_{i,t} = \alpha_i + \beta_i GT_{IT,t} + \delta_i y_{i,t-1} + \gamma_i IV_{i,t} + \omega_i \Delta^{\%}CC_{i,t} + \epsilon_{i,t}, \qquad \epsilon_i \sim \mathcal{N}(0,\sigma_i).
\end{equation*}
Additional robustness checks are available upon request to the authors.

\begin{table}[H] 
\centering 
\small
\resizebox{0.68\textwidth}{!}{
\begin{tabular}{|l|*{6}{c}|} 
\toprule 
& DE& FR& GB& US& IT& ES\\ 
\midrule 
const& 0.006* & 0.003 & 0.001 & 0.006 & 0.005 & 0.009* \\ 
& (0.004) & (0.005) & (0.003) & (0.005) & (0.005) & (0.005) \\ 
$GT_{IT,t}$& -0.048*** & -0.048*** & -0.050*** & -0.061*** & -0.079*** & -0.050*** \\ 
& (0.013) & (0.015) & (0.013) & (0.017) & (0.022) & (0.014) \\ 
$y_{t-1}$& -0.132 & -0.183 & -0.195 & -0.528*** & -0.345** & -0.295* \\ 
& (0.115) & (0.146) & (0.158) & (0.145) & (0.166) & (0.189) \\ 
$IV_t$& 0.029 & 0.026 & 0.030* & 0.025 & 0.043** & -0.001 \\ 
& (0.026) & (0.026) & (0.023) & (0.028) & (0.025) & (0.025) \\ 
$\Delta^{\%} CC_t$& -0.016* & -0.008** & -0.001 & 0.000 & -0.004** & -0.000 \\ 
& (0.010) & (0.005) & (0.006) & (0.001) & (0.002) & (0.010) \\ 
& & & & & & \\ 
\midrule 
$\bar{R}^2$  & 0.137  & 0.151  & 0.107  & 0.281  & 0.201  & 0.162 \\ 
\bottomrule 
\end{tabular}
}
\caption{\footnotesize Estimates of the general model for all countries. Social data: YouTube.\\
{\scriptsize Robust standard errors are in parentheses, *, ** and *** denote statistical significance at 10\%, 5\% and 1\%, respectively.}}
\label{tab:reg_on_IT_vol_cases_YoTu}
\end{table}

\begin{table}[H] 
\centering 
\small
\resizebox{0.68\textwidth}{!}{
\begin{tabular}{|l|*{6}{c}|} 
\toprule 
& DE& FR& GB& US& IT& ES\\ 
\midrule 
const& 0.007* & 0.004 & 0.003 & 0.008** & 0.007* & 0.010** \\ 
& (0.005) & (0.005) & (0.004) & (0.005) & (0.005) & (0.005) \\ 
$GT_{IT,t}$& -0.026** & -0.025** & -0.031*** & -0.037** & -0.049*** & -0.027** \\ 
& (0.012) & (0.013) & (0.013) & (0.020) & (0.017) & (0.015) \\ 
$y_{t-1}$& -0.077 & -0.117 & -0.131 & -0.485*** & -0.244* & -0.245* \\ 
& (0.105) & (0.140) & (0.168) & (0.148) & (0.147) & (0.183) \\ 
$IV_t$& 0.011 & 0.007 & 0.012 & 0.004 & 0.015 & -0.015 \\ 
& (0.029) & (0.027) & (0.023) & (0.027) & (0.028) & (0.025) \\ 
$\Delta^{\%} CC_t$& -0.020** & -0.009* & -0.001 & -0.001 & -0.003 & -0.006 \\ 
& (0.011) & (0.005) & (0.008) & (0.002) & (0.003) & (0.011) \\ 
& & & & & & \\ 
\midrule 
$\bar{R}^2$  & 0.077  & 0.085  & 0.030  & 0.228  & 0.099  & 0.113 \\ 
\bottomrule 
\end{tabular}
}
\caption{\footnotesize Estimates of the general model for all countries. Social data: Google News.\\
{\scriptsize Robust standard errors are in parentheses, *, ** and *** denote statistical significance at 10\%, 5\% and 1\%, respectively.}} 
\label{tab:reg_on_IT_vol_cases_GoNe}
\end{table}

\begin{table}[H] 
\centering 
\small 
\resizebox{0.68\textwidth}{!}{
\begin{tabular}{|l|*{6}{c}|} 
\toprule 
& DE& FR& GB& US& IT& ES\\
\midrule 
const& 0.005 & 0.002 & 0.001 & 0.006 & 0.004 & 0.009** \\ 
& (0.005) & (0.005) & (0.003) & (0.005) & (0.005) & (0.005) \\ 
$GT_{IT,t}$& -0.042*** & -0.047*** & -0.049*** & -0.058*** & -0.081*** & -0.043*** \\ 
& (0.014) & (0.015) & (0.017) & (0.024) & (0.025) & (0.018) \\ 
$y_{t-1}$& -0.097 & -0.158 & -0.161 & -0.501*** & -0.300** & -0.267* \\ 
& (0.109) & (0.140) & (0.167) & (0.149) & (0.149) & (0.186) \\ 
$IV_t$& 0.023 & 0.024 & 0.026 & 0.020 & 0.043** & -0.006 \\ 
& (0.027) & (0.026) & (0.022) & (0.029) & (0.025) & (0.025) \\ 
$\Delta^{\%} CC_t$& -0.018** & -0.008** & -0.002 & -0.000 & -0.004** & -0.005 \\ 
& (0.011) & (0.005) & (0.007) & (0.002) & (0.002) & (0.011) \\ 
& & & & & & \\ 
\midrule 
$\bar{R}^2$  & 0.103  & 0.125  & 0.070  & 0.250  & 0.168  & 0.137 \\ 
\bottomrule 
\end{tabular} 
}
\caption{\footnotesize Estimates of the general model for all countries. Social data: Google Search.\\
{\scriptsize Robust standard errors are in parentheses, *, ** and *** denote statistical significance at 10\%, 5\% and 1\%, respectively.}} 
\label{tab:reg_on_IT_vol_cases_GoSe}
\end{table}

\end{document}